\documentstyle[aps]{revtex} \draft \begin{document}
\title{PERIODIC MODULATION INDUCED INCREASE OF REACTION RATES IN
AUTOCATALYTIC SYSTEMS} \author{T.  Alarc\'on, A.  P\'erez-Madrid, J.M.
Rub\'{\i}} \address{Departament de F\'{\i}sica Fonamental\\ Facultat de
F\'{\i}sica\\ Universitat de Barcelona\\ Diagonal 647, 08028 Barcelona,
Spain\\ }

\maketitle

\begin{abstract} We propose a new mechanism to increase the reactions rates
in multistable autocatalytic systems.  The mechanism is based upon the
possibility for the enhancement of the response of the system due to the
cooperative behavior between the noise and an external periodic modulation.
In order to illustrate this feature we compute the reaction velocities for
the particular case of the Sel'Kov model, showing that they increase
significantly when the periodic modulation is introduced.  This behavior
originates from the existence of a minimum in the mean first passage time,
one of the signatures of stochastic resonance.

\end{abstract}


\section{INTRODUCTION}

In the field of chemical kinetics and in particular in the study of
autocatalytic systems the enhancement of the reaction velocities is a
matter of primary interest.  Some mechanisms, among which we can quote the
chemical reactions assisted by a magnetic field$^{\ref{bib:mag}}$, have
been proposed to this end.  In this paper we propose stochastic resonance
(SR) $^{\ref{bib:benzi}-\ref{bib:vilar}}$ as a new mechanism to increase
the reaction rates and diffusion in autocatalytic systems.  This phenomenon
consists in the enhancement of the response of a system by the noise when
it is under the influence of a periodic input.  Thus, it constitutes a
cooperative phenomenon between noise and an oscillating signal, being
proper to nonlinear dynamical systems, and also to certain class of
nondynamical systems$^{\ref{bib:bezrukov}}$.  Several magnitudes enables
one to characterize the presence of SR.  The most common of them is the
signal-to-noise ratio (SNR), usually used in signal detection problems,
which measures the response of the system to the input.  The existence of
SR is revealed by th presence of a maximum in the SNR for a non-zero noise
level$^{\ref{bib:wiesenfeld}}$.  However, for our purposes we will pay
attention on another signature of the presence of SR in the system:  the
existence of a minimum in the mean first passage time (MFPT) as a function
of the input frequency$^{\ref{bib:zhou}}$.  This quantity is, essentially,
the relaxation time of the system.  Thus, a diminution in the MFPT implies
an increase of the reaction rates.

Although some works have previously been devoted to show the existence of
SR in chemical systems$^{\ref{bib:sch1}-\ref{bib:reichl}}$, the possibility
of improving the reaction rates is an aspect which has not been treated up
to know.  When we are concerned with catalytic systems, the fact that the
MFPT, which can be interperted as the relaxation time, reaches a minimum
for some non-zero noise level leads to an enhancement in the time of
reaction.

Along this paper we will focus our attention on a particular example of
autocatalytic model:  the Sel'Kov model$^{\ref{bib:selkov}}$.  To be
precise, we will use an implementation of this model proposed by Ross
$\it{et\;al.}$.  To achieve this purpose, we will apply to this model a
theory we have developed$^{\ref{bib:nonpot}}$ with the objective of
deriving analytical expressions for the reaction rates and more generally
to analyze the stochastic properties of a certain class of nonpotential
systems, to which the Sel'Kov model belongs.  Making use of this theory we
can extend some previous analysis performed in the framework of potential
systems$^{\ref{bib:reacciones}-\ref{bib:kramers}}$ to the nonpotential case
thus, unifying the treatment for both.

The approach to the problem that we are going to adopt$^{\ref{bib:nonpot}}$
is based on the assumption of weak noise limit and on the particular
structure of the phase space of the dynamical system.  The phase space must
be two dimensional with three aligned fixed points:  one unstable fixed
fixed point between two stable fixed point.  Moreover, one of the variables
has to relax faster than the other, in such a way that we can restrict the
dynamics, in the adiabatic limit, to the line which contains the fixed
points.  The reduction to an one dimensional dynamics, which is obviously
potential, allows us to treat the system in the same way as if it were
potential.

We have organized the paper in the following way.  In Section II we
formulate the kinetic equations for the Sel'Kov model and study the
behavior of the reaction rates under a periodic external forcing.  Section
III is devoted to discuss the stochastic properties of this model.  We show
the existence of SR by computing the signal-to-noise ratio in Section III,
and the distribution of residence times and the MFPT in Section IV.
Finally, in Section V we discuss the implications of our results in
chemical kinetics.

\section{THE SEL'KOV MODEL}

The equations of the Sel'Kov model, as proposed by Ross
$\it{et\;al.}$$^{\ref{bib:ross2}}$ are given by

\begin{equation}\label{eq:a1}
\frac{dp_{X}}{dt}\,=\,k_{1}p_{A}\,+\,k_{4}p_{Y}^{3}\,
-\,k_{2}p_{X}\,-\,k_{3}p_{X}p_{Y}^{2}.  \end{equation}

\begin{equation}\label{eq:a2}
\frac{dp_{Y}}{dt}\,=\,k_{6}p_{B}\,+\,k_{3}p_{X}p_{Y}^{2}\,
-\,k_{5}p_{Y}\,-\,k_{4}p_{Y}^{3}.  \end{equation}

These equations correspond to a reaction occurring isothermally in a region
of constant volume linked to two infinite reservoirs, containing species
$A$ and $B$, by membranes selectively permeable.  This region contains
intermediate species $X$ and $Y$ and the catalysts.  The two reservoirs are
maintained at constant pressures $p_{A}$ and $p_{B}$ whereas in the
intermediate region the pressure is not constant due to the reaction.  Eqs.
(\ref{eq:a1}) and (\ref{eq:a2}) give us the evolution of the vapour
pressures for the intermediate species $P_{X}$ and $P_{Y}$.  The external
periodic forcing will be introduced through periodic variations of $p_{A}$
or $p_{B}$.

\subsection{Kinetic Equations}

Let us briefly discuss the method we have used to obtain the kinetic
equations$^{\ref{bib:nonpot}}$.  The class of two-dimensional nonpotential
noisy dynamical systems to which our method applies is characterized by the
peculiar topology of the phase space.  For the case concerning us, the
dynamics is characterized by the presence of three aligned fixed points:
an unstable fixed points between two stable fixed points.  Moreover, it is
necessary that one of the variables relaxes faster than the other.  In our
case this fast variable is $p_{Y}$ whose relaxation time is, approximately,
$k_{5}^{-1}$ which is smaller than $k_{2}^{-1}$, the corresponding
relaxation time for $p_{X}$$^{\ref{bib:ross2}}$.

Consider a general two-dimensional noisy dynamical
system$^{\ref{bib:maier}}$

\begin{equation}\label{eq:a3}
\frac{d\vec{x}}{dt}\,=\,\vec{u}(\vec{x})\,+\,\vec{\vec{g}}\cdot\vec{\xi}(t),
\end{equation}

\noindent where $\vec{x}$ is the vector whose components are state
variables, the field $\vec{u}(\vec{x})$ is the drift, which is suposed to
fulfill the aforementioned conditions, $\vec{\vec{g}}$ is the noise matrix
and $\xi(t)$ is a gaussian white noise of zero mean and correlation
function given by

\begin{equation}\label{eq:a4} \langle
\xi_{i}(t)\xi_{j}(t')\rangle\,=\,2D\delta_{ij}\delta(t-t'), \end{equation}

\noindent with $D$ accounting for the noise level.

The corresponding Fokker-Planck equation involving the probability density,
$\rho(x,t)$, is

\begin{equation}\label{eq:a5} \frac{\partial\rho}{\partial
t}\,=\,\nabla\cdot(-\vec{u}(\vec{x})\rho+\nabla\cdot(\vec{\vec{D}}\rho)),
\end{equation}

\noindent where $\vec{\vec{D}}\,=\,D\vec{\vec{g}}\cdot\vec{\vec{g}}^{T}$ is
the diffusion tensor, with T indicating transposition.

Let us now assume that the system is potential.  In such a case it is
possible to write the Fokker-Planck equation as a continuity equation

\begin{equation}\label{eq:a6} \frac{\partial\rho}{\partial
t}\,=\,-\nabla\cdot\vec{J}, \end{equation}

\noindent where $\vec{J}$ is the diffusion current given by

\begin{equation}\label{eq:a7} \vec{J}\,=\,-D\,e^{-U/D}\nabla\cdot
e^{\vec{\vec{\mu}}/D}.  \end{equation}

\noindent To obtain this expression we have defined $U$ and
$\vec{\vec{\mu}}$ ( a generalized chemical potential) as follows

\begin{equation}\label{eq:a8} U\,\equiv\,-\frac{1}{2}\int\,\vec{u}\cdot
d\vec{x}, \end{equation}

\begin{equation}\label{eq:a9}
e^{\vec{\vec{\mu}}/D}\,\equiv\,\vec{\vec{g}}\rho e^{U/D}.  \end{equation}

For a potential system the function $U$ defined through (\ref{eq:a8})
coincides with its potential energy.  In the nonpotential case, however,
the value of $U$ will depend on the path of integration we choose and, in
general, we cannot achieve eqs.  (\ref{eq:a6}) and (\ref{eq:a7}).

Let us now assume the adiabatic hypothesis together with the weak noise
limit and analyze the consequences that these assumptions have on the
general properties of the quasi-stationary probability distribution.  If we
allow the system to evolve for sufficiently long time, the probability
distribution will acquire two high maxima around the stable fixed points
(SFP) of the (deterministic) system$^{\ref{bib:ebeling}}$.  In the weak
noise limit, the probability distribution will be very narrow around the
line on which the fixed points lie.  Therefore, we can assume that the
fluctuations are concentrated along this line.  As shown
in$^{\ref{bib:nonpot}}$, the potential for this system is given by the
function $U$ taking precisely this line as integration path.

From eqs.  (\ref{eq:a1}) and (\ref{eq:a2}) by equating their right hand
side to zero and adding the resulting equations, one can see that the fixed
points have to lie on the line

\begin{equation}\label{eq:a10}
p_{Y}\,=\,-\frac{k_{2}}{k_{5}}p_{X}\,+\,\frac{k_{1}p_{A}+k_{6}p_{B}}{k_{5}}.
\end{equation}

\noindent Having in mind that $p_{Y}$ is the fast variable, we can
introduce (\ref{eq:a10}) in (\ref{eq:a1}) to obtain a one dimensional
drift, which, when equated to zero and for the proper values of the
parameters, yields a third order equation with three real different roots,
corresponding to the steady states of the system.  A simple computation
allows us to see that eq.  (\ref{eq:a8}) when integrated along the line
(\ref{eq:a10}) gives the potential for this drift.

Before proceeding further, it is convenient to rescale the resulting one
dimensional system.  To this purpose we will introduce the dimensionless
variables

\begin{equation}\label{eq:a11} x\,=\,\frac{p_{X}}{p_{A}}, \end{equation}

\begin{equation}\label{eq:a12} \tau\,=\,k_{2}t.  \end{equation}

The corresponding Langevin equation describing the dynamics of the
fluctuations along the line defined through eq.  (\ref{eq:a10}) is, in
nondimensional variables,

\begin{equation}\label{eq:a13}
\dot{x}\,=\,a+b\,-\,(1-c)x\,+\,dx^{2}\,-\,ex^{3}\,+\,\xi(\tau),
\end{equation}

\noindent where the coefficients are related to the original parameters by

\begin{eqnarray} a&=&\frac{k_{1}}{k_{2}},\nonumber\\
b&=&K^{3}\frac{k_{4}}{p_{A}},\nonumber\\
c&=&K^{2}(\frac{k_{3}}{k_{4}}+\frac{k_{1}k_{4}}{k_{4}k_{5}}),\\
d&=&K(p_{A}\frac{k_{4}k_{2}}{k_{5}}+2p_{A}\frac{k_{3}}{k_{5}}),\nonumber\\
e&=&p_{A}^{2}\frac{k_{2}}{k_{5}^{2}}(\frac{k_{4}k_{2}}{k_{5}}+k_{3});
\nonumber \end{eqnarray}

\noindent where $K\equiv\frac{k_{1}p_{A}+k_{6}p_{B}}{k_{5}}$.  The
potential as a function of this nondimensional variable is obviously given
by

\begin{equation}\label{eq:a14}
U\,=\,\frac{e}{4}x^{4}-\frac{d}{3}x^{3}+\frac{1+c}{2}x^{2}-(a+b)x.
\end{equation}

In Figs.  1a and 1b we show the fixed points of the Sel'Kov system and the
form of the function $U$, respectively.  From Fig.  1a it is easy to infer
that the unstable fixed point (UFP) is $F_{0}$, while $F_{\pm}$ are the SFP
(of the 2-dimensional model)$^{\ref{bib:selkov}}$.  Fig.  2 shows that the
SFP of the total systems are in this reduced version the minima of the
potential corresponding to the maxima of the probability distribution
function, and the UFP is a maximum of $U$.

Our next step will be to discretize the dynamics on the null cline .  In
particular we will obtain the kinetic equations.  To this end, we define
the populations $n_{+}$ ( $n_{-}$ ) as the population on the right ( left )
of the UFP$^{\ref{bib:gardiner}}$

\begin{equation}\label{eq:a15}
n_{+}\,=\,\int_{\it{S[+]}}\,\rho(\vec{x},t)d\vec{x} \end{equation}

\begin{equation}\label{eq:a16}
n_{-}\,=\,\int_{\it{S[-]}}\,\rho(\vec{x},t)d\vec{x} \end{equation}

\noindent where $\it{S[+]}$ ($\it{S[-]}$) is the portion of the phase space
on the right( left ) of a line passing through the UFP and orthogonal to
the line which contains the fixed points.

In the adiabatic limit, we can assume that the population is strongly
concentrated in a small region around the SFP, as suggested by the picture
of the probability density we have profiled when the maxima in this long
time limit is very high.  Therefore, in this limit we can imagine that the
system reaches a quasi-stationary state in which a quasi-stationary
diffusion current is established.  This current is assumed to be uniform
between the two maxima of the probability density and, in the weak noise
limit concentrates around the line joining the three fixed points

\begin{equation}\label{eq:a17}
\vec{J}(\vec{x},t)\,=\,\vec{J}(x,t)\delta(y-y_{0}(x))\,=\,\vec{J}(t)
\delta(y-y_{0}(x))(\theta(x-x_{+})-\theta(x-x_{-})), \end{equation}

\noindent where $y_{0}(x)$ refers to the line (\ref{eq:a10}) written in
nondimensional variables and $x_{+}$ ( $x_{-}$ ) is the coordinate of the
fixed point on the right ( left ) of the UFP.

The kinetic equation for $n_{+}$ is given by

\begin{equation}\label{eq:a18}
\frac{dn_{+}}{dt}\,=\,\int_{\it{S[+]}}\,\frac{\partial\rho}{\partial
t}d\vec{x}\,=\,-\int_{\it{S[+]}}\,\nabla\cdot \vec{J} d\vec{x}.
\end{equation}

\noindent By using the divergence theorem and the assumptions about the
form of the diffusion current, we have

\begin{equation}\label{eq:a19}
\frac{dn_{+}}{dt}\,=\,\int_{-\infty}^{+\infty}J_{1}(x,t)
\delta(y)dy\,=\,J_{1}(x,t), \end{equation}

\noindent and, proceeding in the same way as for the case of $n_{-}$, we
obtain

\begin{equation}\label{eq:a20} \frac{dn_{-}}{dt}\,=\,-J_{1}(x,t)
\end{equation}

In addition, due to the height of the maxima in the probability density and
the weakness of the noise, we can also consider that equilibrium in each
side of the UFP is reached independently.  Thus, the generalized chemical
potential is given by

\begin{equation}\label{eq:a21}
\mu(\vec{x},t)\,=\,\{\mu(x_{+},t)\theta(x_{0}-x)+
\mu(x_{-},t)\theta(x-x_{0})\}\delta(y-y_{0}(x)), \end{equation}

\noindent where $x_{0}$ is the coordinate of the UFP.  Notice that the
tensorial character of the generalized potential has been removed due to
the reduction to a one dimensional dynamics.  Moreover, we have defined
$\mu\equiv\mu_{11}$.  By using equation (\ref{eq:a21}) in (\ref{eq:a7}) we
then obtain

\begin{equation}\label{eq:a22}
\Psi(\vec{x},t)\,=\,\{\Psi_{+}e^{-(U-U_{+})/D}\theta(x_{0}-
x)+\Psi_{-}e^{-(U- U_{-})/D}\theta(x-x_{0})\}\delta(y), \end{equation}

\noindent where $U$ corresponds to the integral along the adequate path
(\ref{eq:a10}), $U_{+}$ and $U_{-}$ are its values at the SFP ( its minima
), $\Psi\equiv g_{11}\rho$ and $\Psi_{\pm}\equiv g_{11}\rho(x_{\pm},t)$.

In order to obtain the expression for the quasi-stationary current
$J_{1}(t)$ we follow the same procedure as in$^{\ref{bib:reacciones}}$.
Finally, we obtain the kinetic equations for $n_{\pm}$

\begin{equation}\label{eq:a23}
\frac{dn_+}{dt}=-\frac{dn_-}{dt}\,=\,K_{-}n_{-}-K_{+}n_{+}, \end{equation}

\noindent where the kinetic coefficients are given by

\begin{equation}\label{eq:a24} K_{\mp}\,=\,\frac{1}{2\pi}(\vert
U_{0}^{^{\prime\prime}}\vert U_{\mp}^{^{\prime\prime}})^{1/2}
e^{-(U_{0}-U_{\mp})/D}.  \end{equation}

\subsection{Periodically Modulated Sel'kov Model}

Let us now introduce an external periodic forcing in this model by
considering a weak periodic perturbation in $p_{B}$ of the form

\begin{equation}\label{eq:a25} p_{B}\,=\,p_{0}(1+\epsilon(t)),
\end{equation}

\noindent where $\epsilon(t)\equiv\epsilon_{0}\sin(\omega_{0}t)$, with
$\epsilon_{0}$ being a small parameter.  The pressure $p_{A}$ is mantained
constant.  In this case the coefficients $a$ and $e$ remain constant while
$b$, $c$ and $d$ become periodic functions,

\begin{eqnarray} b&=&b_{0}(1+\epsilon(t)),\nonumber\\
c&=&c_{0}(1+\epsilon(t)),\\ d&=&d_{0}(1+\epsilon(t)),\nonumber
\end{eqnarray}

\noindent Since $U$, its derivatives and the positions of the fixed points
are also periodic functions, the reaction rates $K_{\pm}$ oscillate, too.
For these we achieve$^{\ref{bib:nonpot},\ref{bib:wiesenfeld}}$, up to
second order in the parameter $\phi_{0}\equiv\frac{\epsilon_{0}}{D}$ and
under the assumption of weak noise,

\begin{equation}\label{eq:a27}
K_{\pm}(t)\,=\,\alpha_{0}^{\pm}+\alpha_{1}^{\pm}\phi_{0}\sin\omega_{0}t+
\alpha_{2}^{\pm}\phi_{0}^{2}\sin^{2}\omega_{0}t, \end{equation}

\noindent where $\alpha_{0}^{\pm}$, $\alpha_{1}^{\pm}$ and
$\alpha_{2}^{\pm}$ are given by

\begin{equation}\label{eq:a28}
\alpha_{0}^{\pm}\,=\,\frac{e^{-\xi_{0}/D}}{2\pi}f_{0}^{\pm}e^{\xi_{\pm}/D},
\end{equation}

\begin{equation}\label{eq:a29} \alpha_{1}^{\pm}\,=
\,-\frac{e^{-\xi_{0}/D}}{2\pi}
f_{0}^{\pm}e^{\xi_{\pm}/D}(\eta_{0}-\eta_{\pm}), \end{equation}

\begin{equation}\label{eq:a30}
\alpha_{2}^{\pm}\,=\,\frac{1}{2}\frac{e^{-\xi_{0}/D}}{2\pi}
f_{0}^{\pm}e^{\xi_{\pm}/D}(\eta_{0}-\eta_{\pm})^{2}.  \end{equation}

\noindent Here $f_{0}^{\pm}$ is the zero order contribution to $(\vert
U_{0}^{\prime\prime}\vert U_{\pm}^{\prime\prime})^{1/2}$.  Notice that the
existence of a periodic forcing is responsible for the time dependence of
the reaction rates.

In Fig.  2a we have plotted $K_{+}$ as a function of time for a fixed noise
level.  As we can see from this figure, there appear periodic oscillations
in the reaction rates.  This behavior is a consequence of the form of the
distribution of residence times ( see Section IV).  An interesting
phenomenon occurs when the noise level is very low.  In this case, the
reaction velocities become very narrow around the maxima and are
practically zero at any other time.

Let us now focus our attention on the time averages of the reaction rates.
These are given by

\begin{equation}\label{eq:a31} \langle
K_{\pm}\rangle_{t}\,=\,\frac{\omega_{0}}{2\pi}
\int_{0}^{\frac{2\pi}{\omega_{0}}}dtK_{\pm}(t)\,=\,
\frac{f_{0}^{\pm}}{2\pi}e^{(-\xi_{0}/D)}(1+\frac{1}{4}
\phi_{0}^{2}(\eta_{0}-\eta_{\pm})^{2}) \end{equation}

\noindent It is important to note that, although in some time intervals the
reaction rates are smaller than in the non perturbed case, the average over
a period of the external forcing is greater than the value of the reaction
rate in the non perturbed case, corresponding to the same noise level.
This can be seen by comparison of Figs.  2b and Fig 2c.

\section{POWER SPECTRUM}

We are now in a position to obtain the stochastic properties of the Sel'Kov
model.  In particular, we will compute the power spectrum and from it the
signal-to-noise ratio (SNR).

In order to compute these quantities, we assume that, in the limit of weak
noise, the probability density ( in one dimension ) can be written
as$^{\ref{bib:wiesenfeld}}$

\begin{equation}\label{eq:c14}
p(v,t)\,=\,n_{+}(t)\delta_{v,v_{+}}+n_{-}(t)\delta_{v,v_{-}},
\end{equation}

\noindent where $n_{+}$ and $n_{-}$ evolve in time according to the kinetic
equations (\ref{eq:a23}).  The formal solution of these equations is found
to be

\begin{equation}\label{eq:c15}
n_{\pm}(t)\,=\,g^{-1}(t)(n_{\pm}(t_{0})g(t_{0})+\int_{t_{0}}^{t}\,
K_{\mp}(t')g(t')dt'), \end{equation}

\noindent with

\begin{equation}\label{eq:c16} g(t)\,=\,\int^{t}(K_{+}+K_{-})dt'.
\end{equation}

Introducing the expansion (\ref{eq:a27}) for the reaction rates in
(\ref{eq:c15}) and (\ref{eq:c16}) one obtains, up to first order in
$\phi_{0}$,

\begin{eqnarray} n_{\pm}(t)\,
&=&\,e^{-\alpha_{0}(t-t_{0})}(\delta_{v(t_{0}),v_{\pm}}-
\frac{\alpha_{0}^{\mp}}{\alpha_{0}}-\phi_{0}\frac{\alpha_{0}^{\mp}
\alpha_{1}}{\alpha_{0}\omega_{0}}cos(\omega_{0}t)+\phi_{0}
\alpha_{1}^{\mp}\frac{cos(\omega_{0}t_{0}+\Phi)}{(\alpha_{0}^{2}
+\omega_{0}^{2})^{1/2}}-\nonumber\\ &
&\phi_{0}\frac{\alpha_{1}}{\omega_{0}}
\frac{cos(\omega_{0}t_{0}+\Phi)}{(\alpha_{0}^{2}+
\omega_{0}^{2})^{1/2}})+\frac{\alpha_{0}^{\mp}}{\alpha_{0}}
(1+\phi_{0}\frac{\alpha_{1}}{\omega_{0}}cos(\omega_{0}t))-
\phi_{0}\alpha_{1}^{\mp}\frac{cos(\omega_{0}t+\Phi)}{(\alpha_{0}^{2}
+\omega_{0}^{2})^{1/2}}+\nonumber \\ &
&\phi_{0}\frac{\alpha_{1}}{\omega_{0}}
\frac{sin(\omega_{0}t+\Phi)}{(\alpha_{0}^{2}+\omega_{0}^{2})^{1/2}},
\end{eqnarray}

\noindent where $\Phi\equiv arctg(\alpha_{0}/\omega_{0})$.  The quantity
$n_{\pm}(t\vert v_{t_{0}},t_{0})$ is the conditional probability that
$v(t)$ is in the + state at time $t$, given that the state at time $t_{0}$
was $v_{t_{0}}$.  From this equation it is possible to compute the
statistical properties of the process $v(t)$.  Of particular interest to
our purposes is to find its autocorrelation
function$^{\ref{bib:wiesenfeld}}$

\begin{equation}\label{eq:c20} \langle
v(t)v(t+\tau)\rangle\,=\,lim_{t\rightarrow -\infty}\langle
v(t)v(t+\tau)\vert v_{t_{0}},t_{0} \rangle.  \end{equation}

\noindent The conditional correlation function is given by

\begin{eqnarray} \langle v(t)v(t+\tau)\vert v_{t_{0}},t_{0}\rangle
&=&v_{+}(t+\tau)v_{+}(t)n_{+}(t+\tau\vert v_{+},t)n_{+} (t\vert
v_{t_{0}},t_{0})+\nonumber\\ & &v_{+}(t+\tau)v_{-}(t)n_{+}(t+\tau\vert
v_{-},t)n_{-}(t\vert v_{t_{0}},t_{0})+\nonumber\\ &
&v_{-}(t+\tau)v_{+}(t)n_{-}(t+\tau\vert v_{+},t)n_{+}(t\vert
v_{t_{0}},t_{0})+ \nonumber \\ & &v_{-}(t+\tau)v_{-}(t)n_{-}(t+\tau\vert
v_{+},t)n_{-}(t\vert v_{t_{0}},t_{0}).  \end{eqnarray}

Let us make some considerations which will allows us to simplify the
computation of the autocorrelation function.  It is clear from equation
(38) that the Fourier transform of the autocorrelation function will depend
on $t$ as well as on the frequency.  This dependency is avoided by taking
its average over the period of the external
forcing$^{\ref{bib:wiesenfeld}}$.  The autocorrelation function is then
computed up to second order in the parameter $\phi_{0}\sim D^{-1}$.Thus, in
the limit of the weak noise $D^{-1}$ it can be neglected when compared with
$D^{-2}$.  Therefore, the only contribution of the first order term of
$v_{\pm}$ to the autocorrelation function comes from its product with the
zero order term of the product of $n$'s.  But on performing the average
this term vanishes.  Moreover, the position of $F_{-}$ is very near to
zero.  In fact, it is an order of magnitude smaller than $\epsilon_{0}$ and
therefore it can be neglected.  Finally one arrives at

\begin{equation}\label{eq:c21} \overline{\langle v(t)v(t+\tau)\vert
v_{t_{0}},t_{0}\rangle}\,=\,\eta_{+}^{2} \overline{n_{+}(t+\tau\vert
v_{+},t)n_{+}(t\vert v_{t_{0}},t_{0})}, \end{equation}

\noindent where the bar indicates average over $t$.  From equations (38)
and (\ref{eq:c21}) taking the average and the limit $t_{0}\rightarrow
-\infty$ we finally obtain the following expression for the autocorrelation
function

\begin{eqnarray} \overline{\langle v(t)v(t+\tau)\rangle}\,
&=&\,\eta_{+}^{2}
(\frac{\alpha_{0}^{-}}{\alpha_{0}})^{2}+e^{-\alpha_{0}\vert\tau\vert}
(\frac{\alpha_{0}^{-}}{\alpha_{0}})(1 -\frac{\alpha_{0}^{-}}{\alpha_{0}})+
\nonumber \\ &
&\eta_{+}^{2}\phi_{0}^{2}\{\frac{1}{2}e^{-\alpha_{0}\vert\tau\vert}
(\frac{\alpha_{0}^{-}\alpha_{1}^{-}\alpha_{1}}{\alpha_{0}}-\alpha_{0}^{-}
(\frac{\alpha_{1}}{\omega_{0}})^{2}-(\alpha_{1})^{2}-
(\frac{\alpha_{1}}{\omega_{0}})^{2})(\frac{1}{\alpha_{0}^{2}+
\omega_{0}^{2}})+ \nonumber \\ &
&(\frac{\alpha_{0}^{-}\alpha_{1}^{-}\alpha_{1}}{\alpha_{0}}-\alpha_{0}
(\frac{\alpha_{1}}{\omega_{0}})^{2})\frac{cos(\omega_{0}\tau)}{\alpha_{0}
^{2}+\omega_{0}^{2}}+ \nonumber \\ &
&(\frac{\alpha_{0}^{-}}{\alpha_{0}}\frac{\alpha_{1}^{2}}{\omega_{0}}
+\frac{\alpha_{0}^{-}\alpha_{1}^{-}\alpha_{1}}{\alpha_{0}})
sin(\omega_{0}\tau))+ \nonumber \\ &
&\frac{cos(\omega_{0}\tau)}{\alpha_{0}^{2}+\omega_{0}^{2}}
(\frac{\alpha_{0}^{-}\alpha_{1}^{2}}{\omega_{0}^{2}}-
\frac{\alpha_{0}^{-}\alpha_{1}^{-}\alpha_{1}}{\alpha_{0}}+
\frac{1}{2}(\alpha_{1}^{-})^{2}+\frac{1}{2}
(\frac{\alpha_{1}}{\omega_{0}})^{2})\}.  \end{eqnarray}

Whith these results we can now compute the average of the power spectrum
defined as

\begin{equation}\label{eq:c23}
\overline{S(\Omega)}\,=\,\frac{\omega_{0}}{2\pi}\int_{0}^{2\pi/\omega_{0}}
S(\Omega ,t)dt\,=\,\int_{-\infty}^{+\infty}\overline{\langle
v(t)v(t+\tau)\rangle}e^{-\imath\Omega\tau}d\tau, \end{equation}

\noindent where the last equality follows from the commutative character of
the average and Fourier transform.  After Fourier transforming (40), we
obtain

\begin{eqnarray} \overline{S(\Omega)}\,
&=&\,\eta_{+}^{2}\frac{\alpha_{0}^{-}}{\alpha_{0}}
\delta(\Omega)+2\frac{\eta_{+}^{2}}{\alpha_{0}^{2}+
\Omega^{2}}\alpha_{0}^{-}(1-\frac{\alpha_{0}^{-}}{\alpha_{0}})
+2\pi\frac{\eta_{+}^{2}\phi_{0}^{2}}{\alpha_{0}^{2}+\Omega^{2}}
(\frac{(\alpha_{0}^{-})^{2}\alpha_{1}^{2}}{\omega_{0}^{2}}- \nonumber \\ &
&\frac{(\alpha_{0}^{-})^{2}\alpha_{1}^{-}\alpha_{1}}{\alpha_{0}^{2}}+
\frac{1}{2}(\frac{\alpha_{1}^{-1})^{2}}{\alpha_{0}^{-}}+\frac{1}{2}
(\frac{\alpha_{1}\alpha_{0}^{-}}{\omega_{0}})^{2})
\delta(\Omega-\omega_{0}).  \end{eqnarray}

\noindent In this expression, the fraction of the total power in the
broadband noisy part of the spectrum, which usually is a small fraction of
the total power, has been neglected$^{\ref{bib:wiesenfeld}}$.  Note that
the power spectrum contains a term proportional to $\delta(\Omega)$.  The
appearance of this term is due to the asymmetry of the potential that
originates a mean probability current between the two stable states, i.e.
towards the deepest well of the potential.  This corresponds to the
(deterministic) approach to the equilibrium state.

From equation (42) the signal to noise ratio, SNR, can be obtained as a
function of the noise level $D$, by making $\Omega=\omega_{0}$.

\begin{equation}\label{eq:c25}
SNR\,=\,\frac{\frac{\epsilon_{0}^{2}\pi}{D^2}(\frac{3}{2}
\alpha_{0}^{-}(\frac{\alpha_{1}}{\omega_{0}})^{2}-
\frac{\alpha_{0}^{-}\alpha_{1}^{-}\alpha_{1}}{\alpha_{0}^{2}}
+\frac{1}{2}\frac{(\alpha_{1}^{-})^{2}}{\alpha_{0}^{-}})}{1-
\frac{\alpha_{0}^{-}}{\alpha_{0}}}.  \end{equation}

This quantity has been plotted in Fig.  3 as a function of the noise level
or input noise.  It exhibits a maximum at a certain value of $D$, thus
indicating the existence of stochastic resonance.

\section{ESCAPE-TIME DISTRIBUTION AND MEAN FIRST PASSAGE TIME}

In order to compute the escape-time distribution ( ETD ) around the fixed
points of the system we will assume the existence of an absorbing barrier
between them$^{\ref{bib:zhou}}$.  This implies that the kinetic equations
reduce to

\begin{equation}\label{eq:d1} \frac{dn_{\pm}}{dt}\,=\,-K_{\pm}n_{\pm}.
\end{equation}

In this case, and using initial conditions $n_{\pm}(t=0)=1$, the solution
of eq.  (\ref{eq:d1}) can be written as

\begin{equation}\label{eq:d2}
n_{\pm}(t)\,=\,\exp\{-\frac{1}{\omega_{0}}\int_{0}^{\omega_{0}t}\,dzK_{\pm}(z)\}.
\end{equation}

\noindent On the other hand, the ETD around the fixed point
$F_{\pm}$$^{\ref{bib:zhou}}$ is given by

\begin{equation}\label{eq:d3} \rho_{\pm}\,=\,-\frac{dn_{\pm}}{dt}.
\end{equation}

\noindent Therefore, the ETD reads

\begin{equation}\label{eq:d4}
\rho_{\pm}\,=\,K_{\pm}\exp\{-\frac{1}{\omega_{0}}
\int_{0}^{\omega_{0}t}\,dzK_{\pm}(z)\}.  \end{equation}

\noindent Performing an expansion up to second order in the parameter
$\phi_{0}$ we obtain

\begin{eqnarray}\label{eq:d5}
\rho_{\pm}&=&(\alpha_{0}^{\pm}+\alpha_{1}^{\pm}\phi_{0}\sin\omega_{0}t+
\alpha_{2}^{\pm}\phi_{0}^{2}\sin^{2}\omega_{0}t)\times\nonumber \\ &
&\exp\{-\alpha_{0}^{\pm}t+
\frac{\alpha_{1}^{\pm}}{\omega_{0}}\phi_{0}(1-\cos\omega_{0}t)+
\frac{\alpha_{2}^{\pm}}{\omega_{0}}\phi_{0}^{2}(\frac{\sin
2\omega_{0}t}{4}-\frac{\omega_{0}t}{2})\}.  \end{eqnarray}

\noindent From this equation we can see that a coherent response of the
system is generated by the input signal in the sense that the distribution
is modulated by the signal.  We have plotted this quantity in Fig.  4.

As follows from the definition of the ETD, the quantity $\rho_{\pm}dt$
corresponds to the probability that the absorbing boundary to be reached when
the evolution of the system comes from the fixed point $F_{\pm}$.  The MFPT
of leaving $F_{+}$ is given by

\begin{equation}\label{eq:d6} \langle
T\rangle\,=\,\int_{0}^{\infty}dt\,t\rho_{+}(t)\,=
\,\int_{0}^{\infty}dtn_{+}(t).  \end{equation}

\noindent where use has been made of eq.  (\ref{eq:d3}) and an integration
by parts has been performed.  Using eq.  (\ref{eq:d2}) with eq.
(\ref{eq:a27}) we obtain

\begin{equation}\label{eq:d7} \langle
T\rangle\,=\,\int_{0}^{\infty}dt\exp\{-\frac{1}{\omega_{0}}(\alpha_{0}^{+}t+
\alpha_{1}^{+}\phi_{0}(1-\cos\omega_{0}t)+
\alpha_{2}^{+}\phi_{0}^{2}(\frac{\omega_{0}t}{2}-\frac{\sin
2\omega_{0}t}{4})\}.  \end{equation}

\noindent Taking into account that $\epsilon_{0}$ is a small parameter, we
can perform a Taylor expansion of the exponential.Up to second order we
obtain

\begin{eqnarray}\label{eq:d8} \frac{\langle T\rangle}{\langle
T_{0}\rangle}&=&1-
\alpha_{1}^{+}\phi_{0}\frac{\omega_{0}}{(\alpha_{0}^{+})^{2}+
\omega_{0}^{2}}\nonumber\\ &
&+\frac{1}{2}({\alpha_{1}^{+}\phi_{0})^{2}\frac{6\omega_{0}^{2}}{((\alpha_{0}^{+})^2
+4\omega_{0}^{2})((\alpha_{0}^{+})^{2}+\omega_{0}^{2})}}\nonumber\\ &
&-\frac{1}{2}\frac{\alpha_{2}^{+}\phi_{0}^{2}}{\alpha_{0}^{+}}
\frac{4\omega_{0}^{2}}{(\alpha_{0}^{+})^{2}+\omega_{0}^{2}}.
\end{eqnarray}

\noindent where $\langle T_{0}\rangle$ is the MFPT in the absence of
external forcing.  We have plotted this quantity in Fig.  5.

\section{DISCUSSION}

In this paper we have proposed a new mechanism to increase the reaction
rates or equivalently to reduce the reaction times in multistable
autocatalytic systems.  This mechanism is based upon the existence of SR in
this kind of systems, which consists essentially in the enhancement of the
response of the system to a periodic input and the diminution of the MFPT.

We have found that the introduction of a weak periodic forcing improves the
reaction velocities.  As concluded from Fig.  2b, 2c and 3 the values of
the input noise for which this effect is more important correspond,
roughly, to the interval of noise intensity where SR occurs.  To be
precise, for $D=0.19$, corresponding to the appearance of SR in the
system, there is a significant increase in the reaction rate of about 40
per cent.

Another interesting result about the behavior of the reaction rates
concerns its temporal dependence, which can be explained on the light of
the behavior of the ETD.  Comparing Figs.  2a and 4, one can observe that
the reaction rate reaches its maximum value when the corresponding ETD is
at a maximum and vice versa.  hence, when this probability reaches a
maximum value the corresponding transition rate must also be maximum.

With regards to the time dependence of the reaction velocities, we have
found that, in the limit of very weak noise, they become very narrow around
the maxima and near zero for any other time.  In order to explain this fact
let us come back to the picture of the quasi-stationary probability
density, solution of the Fokker-Planck equation, described in section II.
The width of the oscillations around the maxima in the reaction rates, as
well as in the ETD, are due to the transitions induced by the noise.  In a
situation of very low noise intensity these transitions are practically
absent.  In this case the transitions between the fixed points have their
origin, mainly, in the deterministic dynamics which allows the system to go
from one fixed point to another, as reflected in (42) by the presence of
the term proportional to $\delta(\Omega)$, and in the periodic modulation of
the probability density.  This makes the height of the maxima of the
probability density to change along the time.  Consequently, the higher of
them becomes shorter than the other one and this produces a current to the
new most probably position.  This last effect is responsible for the
periodicity observed in the transition rates, whereas the 
transitions between peaks originate from the peculiar deterministic dynamics.

It is important to note that these features about the behavior of the
probability density are reproduced by the function $U$.  For instance, the
aforementioned periodicity of the transition rates can also be understood
in terms of the periodic modulation of the function $U$, which plays the
role of a potential.

On the other hand, due to the existence of SR in the system, the MFPT,
which can be resembled to the reaction time, exhibits a minimum as a
function of the frequency of the periodic input.  Moreover, in the context
of transport phenomena a MFPT can be understood as a relaxation time which,
in essence, is the inverse of a diffusion coefficient.  Therefore, a
minimum in MFPT implies an improvement in the diffusion of the system.  As
one can see from Fig.  5 the diminution in the MFPT with respect to the
case in which the system is not under the action of a periodic forcing is
significantly about 20-25 per cent.

\acknowledgments

This work has been supported by DGICYT of the Spanish Government under
grant PB95-0881.  One of us (T.  Alarc\'on) wishes to thank to DGICYT of
the Spanish Government for financial support.

\newpage

\vspace{3cm} \large

\begin{center} FIGURE CAPTIONS \end{center} \begin{itemize}

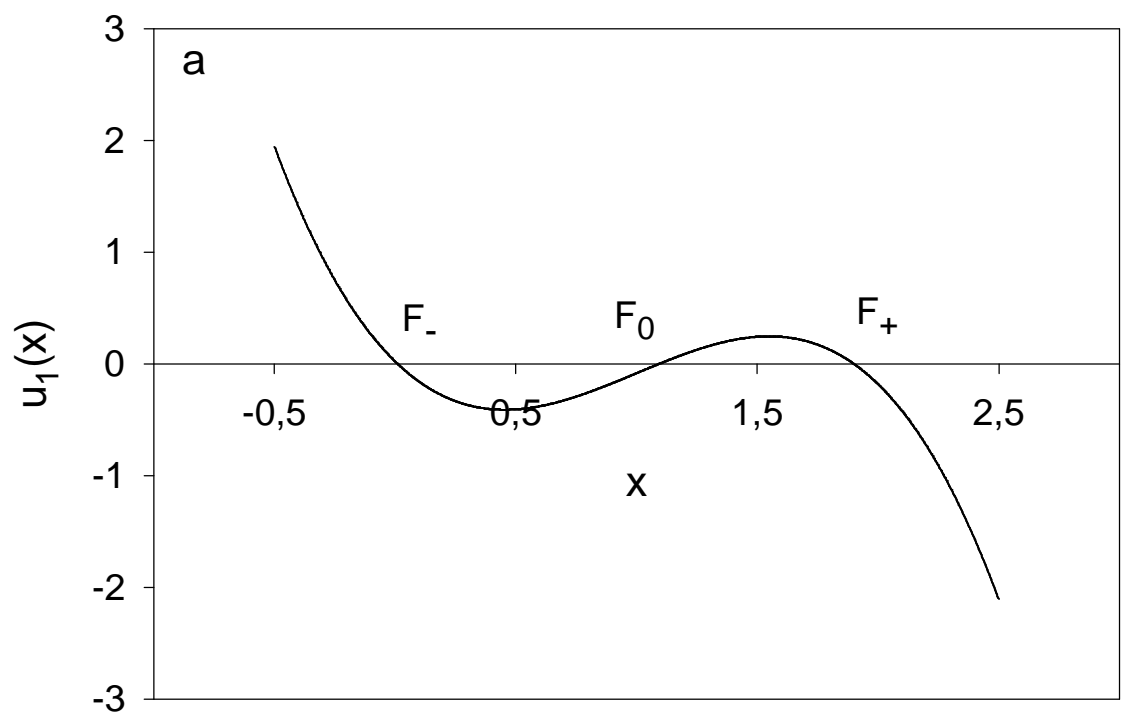
\item{Figure 1}.- \begin{itemize}\item{(a)} Fixed points of the Sel'Kov
model for $a=b_{0}=0.5$, $c_{0}=1.1$, $d_{0}=3$ and $e=1$.  \item{(b)}
Asymmetric one dimensional potential as a function of $x$.\end{itemize}

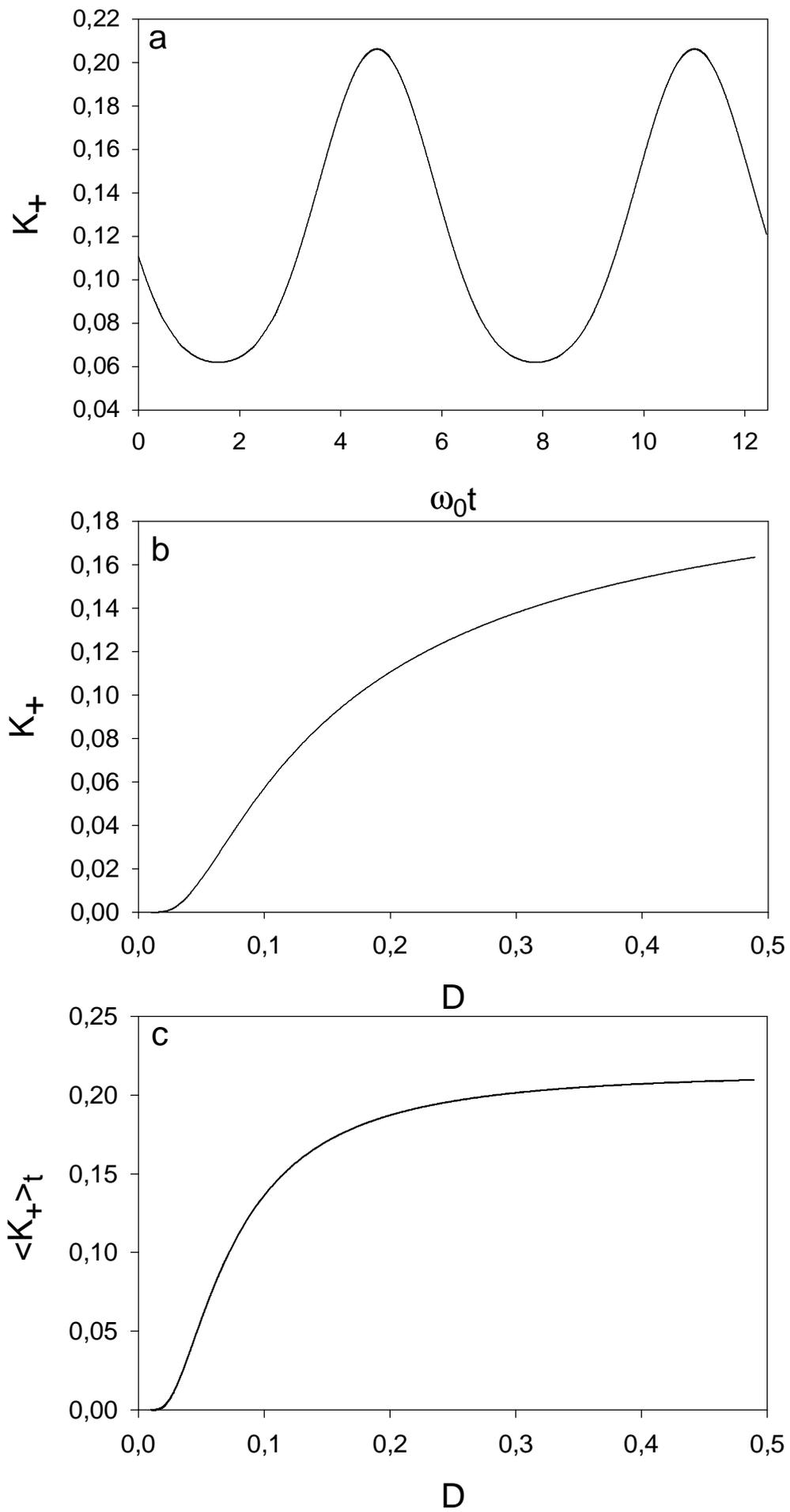
\item{Figure 2}.- \begin{itemize}\item{(a)} $K_{+}$ as a function of time
for $D=0.19$ and $\epsilon_{0}=0.03$.  \item{(b)} $K_{+}$ in absence of
periodic modulation as a function of $D$.  \item{(c)} Temporal average of
$K_{+}(t)$ as a function of $D$ for $\epsilon_{0}=0.03$.\end{itemize}

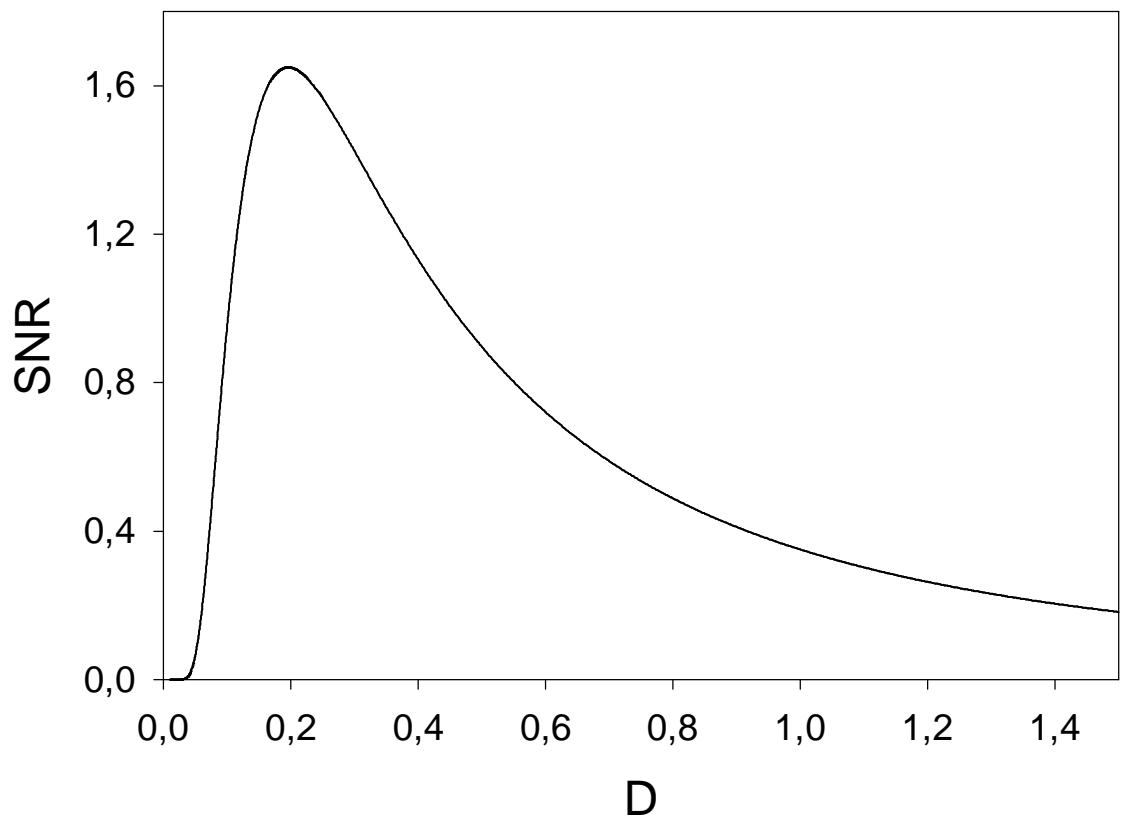
\item{Figure 3}.- Signal-to-Noise Ratio as a function of the noise level
$D$ for $\epsilon_{0}=0.03$.

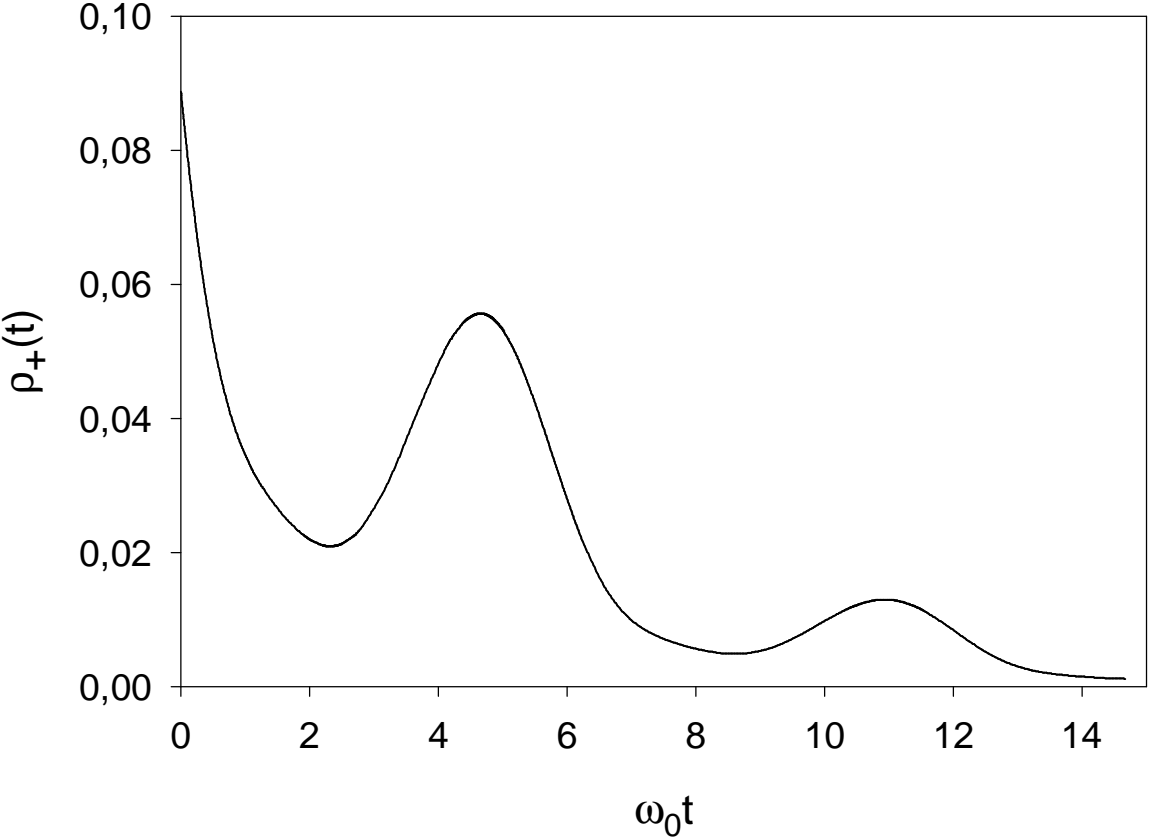
\item{Figure 4}.- Escape time distribution as function of $\omega_{0}t$ for
$D=0.19$ and $\epsilon_{0}=0.03$.

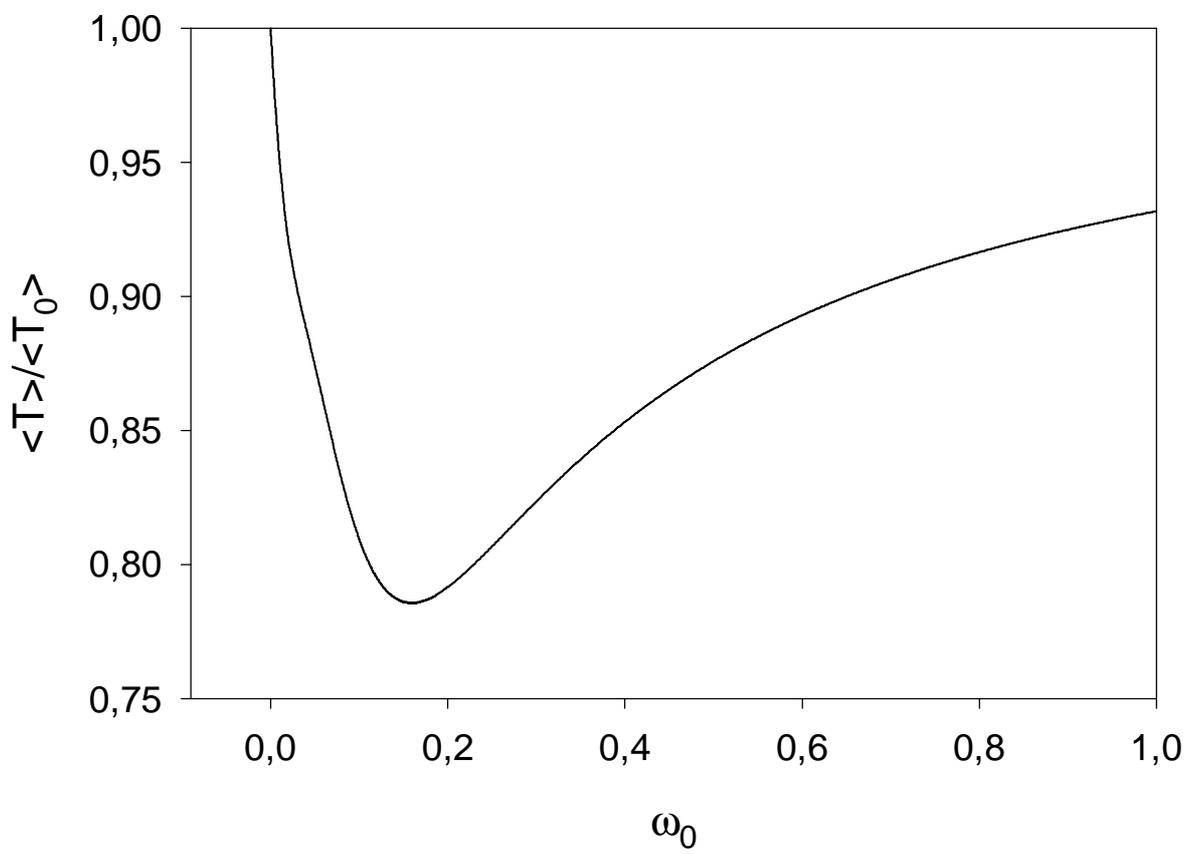
\item{Figure 5}.- Mean first passage time as a function of $\omega_{0}$ for
$D=0.19$ and $\epsilon_{0}=0.03$.

\end{itemize}

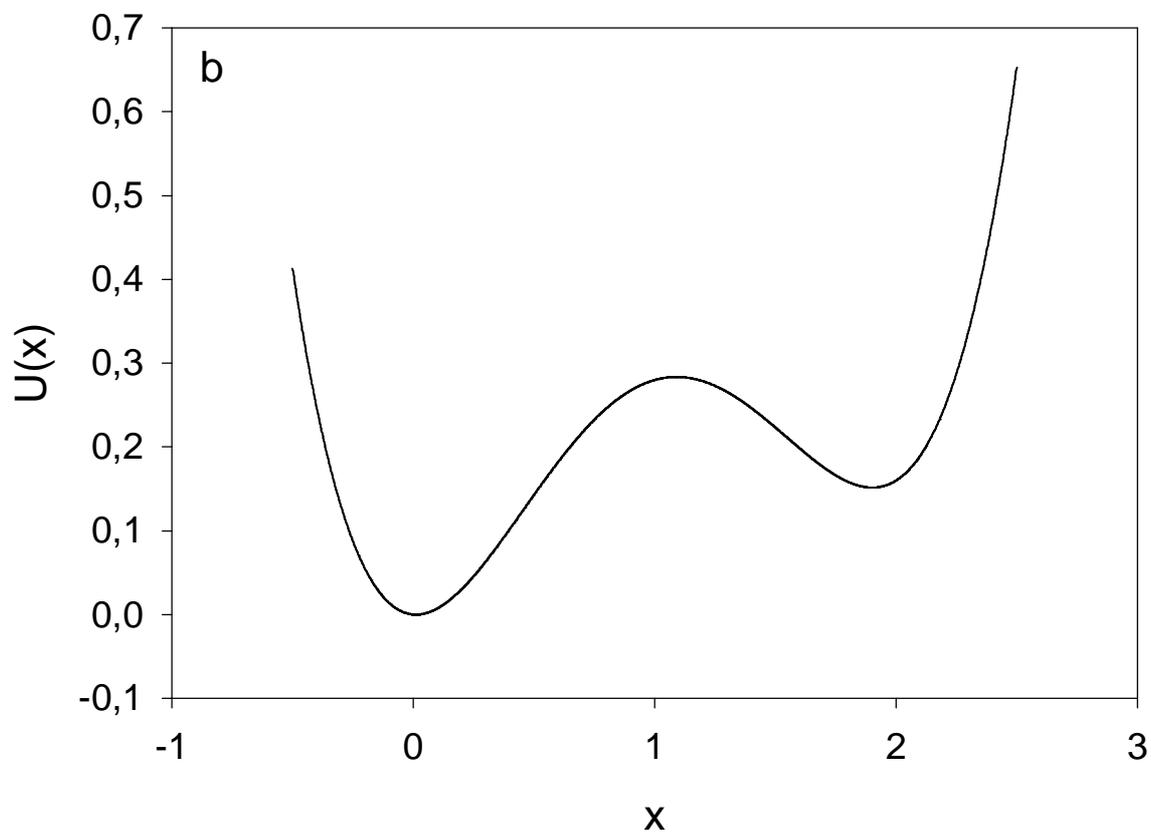
\end{document}